\documentclass[a4paper,10pt,twoside]{cpc-hepnp}

\usepackage{multicol}
\usepackage{graphicx}
\usepackage{amssymb,bm,mathrsfs,bbm,amscd}
\usepackage[tbtags]{amsmath}
\usepackage{lastpage}

\newcommand{\ee}{e^{+} e^{-}}

\newcommand{\jp}{J/\psi}
\newcommand{\psip}{\psi '}

\newcommand{\etac}{\eta_c}
\newcommand{\etacp}{\eta^{\prime}_c}
\newcommand{\mumu}{\mu^{+}\mu^{-}}
\newcommand{\pipi}{\pi^{+}\pi^{-}}

\newcommand{\rt}{\rightarrow}
\newcommand{\etal}{{\em et al.}}

\begin{document}

\title{What's new with the $XYZ$ mesons?
\thanks{Talk given at the Belle-BES-CLEO-BaBar Joint Workshop on Charm 
Physics, Beijing, China November 26-27, 2007}
}

\author{%
Stephen L. OLSEN\email{solsen@ihep.ac.cn}
}

\maketitle

\address{
Institute of High Energy Physics, Chinese Academy of Sciences, Beijing 
100049, China~ \&  \\
Department of Physics \& Astronomy, University of Hawaii, 
Honolulu, HI 96822 
USA}

\begin{abstract}
I review some of the recent experimental results on the
so-called $XYZ$ mesons. 
\end{abstract}

\begin{keyword}
charmonium, multiquark mesons, quark-gluon hybrids
\end{keyword}

\begin{pacs}
14.40.Gx, 12.39.Mk, 13.25.Hw
\end{pacs}

\footnotetext[0]{\hspace*{-2em}\small\centerline{\thepage\ --- 
\pageref{LastPage}}}%

\begin{multicols}{2}

\section{Introduction}

In addition to carrying out their day jobs of studying $CP$ violation
and measuring CKM matrix elements, the $B$-factory experiments have
discovered a number of interesting charmonium-like meson states that
are known collectively as the ``$XYZ$'' mesons.  
Two of these have been given assignments as
charmonium states: the $\eta_c^{\prime}$,\cite{skchoi_etac2s}  
and the $\chi_{c2}'$.\cite{belle_z3940}  However,
the others have properties that are at odds with 
expectations of the  charmonium model and, as a result,
remain unclassified.  These
latter include the $X(3872)$\cite{belle_x3872} and  
the $Y(4260)$,\cite{babar_y4260} which decay to 
$\pipi\jp$; the $X(3940)$,\cite{belle_x3940} seen in $D^*\bar{D}$; the
$Y(3940)$,\cite{belle_y3940} seen in $\omega\jp$;
and the $Y(4325)$,\cite{babar_y4325} seen in $\pi^+\pi^- \psip$.

Proposed assignments for these states have included:
multiquark states, either of the $(c\bar{q},\bar{c}q)$ ``molecular''
type\cite{molecular} or $[cq,\bar{c}\bar{q}]$ diquark-antidiquark
type\cite{diquark} (here $c$ represents a charmed quark and $q$
either a $u$-, $d$- or $s$-quark); hybrid $c\bar{c}$-gluon 
mesons;\cite{hybrid}
or other missing charmonium states where the masses 
predicted by potential models are 
drastically modified by nearby $D^{(*)}\bar{D^{(*)}}$ 
thresholds.\cite{eichten, chao}
A characteristic that would clearly distinguish a multiquark 
state from hybrids or charmonia is the
possibility to have  mesons with non-zero
charge (e.g. $[cu\bar{c}\bar{d}]$), strangeness
($[cd\bar{c}\bar{s}]$)  or both ($[cu\bar{c}\bar{s}]$).\cite{chiu}

During the past summer a large number of new results related to
the $XYZ$ mesons has been reported, which I will try to summarize in 
this talk. My focus will be almost entirely experimental.

\section{What's old?}

First, I briefly summarize  the status before Summer 2007.

\paragraph{$X(3872)$~~~~}Measurements at CDF\cite{CDF_JPC} and 
Belle\cite{Belle_JPC} have favored a $J^{PC}=1^{++}$ assignment for 
the $X(3872)$, although $2^{-+}$ cannot be completely ruled out.  The
two charmonium assignments that match these quantum numbers are
the $\chi^{\prime}_{c1}$ ($2^3P_1$) and the $\eta_{c2}$ ($1^1D_2$).
The mass is too low for the $\chi^{\prime}_{c1}$, especially
if Belle's $\chi^{\prime}_{c2}$ 
candidate (with $M=3931$~MeV) has been
correctly assigned, and too high for the 
$\eta_{c2}$.\cite{barnes:2005,quigg:2006}. In addition
for either assignment, the decay to the $\pipi\jp$ ``discovery
mode'', which BaBar has shown 
has a branching fraction that is above~4\%,\cite{BaBar_B2jpsi-incl}  
is  isospin violating, and should be suppressed.  Thus,
the consensus opinion is that there is no acceptable charmonium
assignment for the $X(3872)$, although this is not a unanimously
accepted point of view.\cite{chao}

\paragraph{$X(3940)$~~~~}  The $X(3940)$ 
was seen by Belle\cite{belle_x3940}
recoiling from the $\jp$ in the $\ee$ continuum annihilation 
process $ee \rt \jp D \bar{D^*}$. (In this report, the inclusion
of the charge conjugate mode is always implied.)
Since the only known charmonium
states that are produced this way are the $0^{-+}$ $\etac$ and 
$\etacp$, and the $0^{++}$ $\chi_{c0}$, circumstantial evidence
suggests that the $X(3940)$ is either a scalar or pseudoscalar. The
lack of evidence for a $D\bar{D}$ decay mode favors the
pseudoscalar $0^{-+}$ assignment.  The possible charmonium 
assignment is the $\eta_c^{\prime\prime}$ ($3^1S_0$).
This has some difficulty because the $3^3S_1$ state is
the $\psi(4040)$ and its mass is pretty well established
at 4040~MeV.  So, an $\eta_c^{\prime\prime}$ assignment for the $X(3940)$
implies a  singlet-triplet mass splitting for radial quantum number
$n=3$ ($\simeq 100$~MeV) that is larger than that for 
$n=2$ ($\simeq 50$~MeV).  Eichten, Quigg and Lane\cite{quigg:2006}
have shown that this may be due to large admixtures
of $D\bar{D^*}$ and $D^*\bar{D^*}$ components in the
$\eta_c^{\prime\prime}$ and $\psi(4040)$ wave functions. 

\paragraph{$Y(3940)$~~~~}  Not much is known about the
$Y(3940)$ other than it must have $C=+$.  If
we assume that the branching fraction for $B\rt KY(3940)$
is not larger than $10^{-3}$, which is typical for
factorization-allowed decay modes such as
$B\rt K \jp)$ and $B\rt K\etac$, the measured product
branching fraction  
${\cal B}(B\rt K Y)\times{\cal B}(Y\rt \omega\jp)=7\pm 3 \times 10^{-5}$
and measured width $\Gamma= (87 \pm34)$~MeV imply  a
large value for the partial width
$\Gamma(Y(3940)\rt \omega \jp)\sim$few~MeV.
This is much larger than those seen for hadronic transitions
between established charmonium states, which are at most $\sim 100$~keV.
Although the
$X(3940)$ (discussed above) is not seen to decay to $\omega\jp$  the
upper limit on this branching fraction ($\le 26\%$ at 90\% CL)
is not stringent enough to rule out the possibility that
it and the $Y(3940)$ are the same states.  If so, the
large $\Gamma(Y(3940)\rt \omega \jp)$ partial width
raises problems with the $\eta_c^{\prime\prime}$ assignment.

\paragraph{$X(4260)$~~~~}
The $Y(4260)$ was discovered by BaBar 
in a 223~fb$^{-1}$ data sample as a 
relatively narrow ($\Gamma = 88\pm 24$~MeV) peak 
near 4260~MeV in the $\pipi\jp$
invariant mass distribution in  the reaction
$e^+e^-\rt \gamma\pipi\jp$, where the $\gamma$ 
exhibits the distinct angular behavior of 
initial state radiation.\cite{babar_y4260}
Since it is clear that this state are produced by 
radiative-return $s$-channel $e^+e^-$ annihilation, it must have 
$J^{PC}=1^{--}$.   In this case, the $Y(4260)$ decaying into 
pairs of open-charm mesons might be seen in the 
total cross section 
$e^+e^-\rt$~hadrons near $E_{cm}=4260$~MeV. However, this
has been measured rather precisely in small $E_{cm}$ bins
$(\sim 10$~MeV) over this energy region by both
the BES\cite{bes_2003R} and Crystal Ball\cite{CB_R}
experiments, and neither group sees evidence for structure
in this region.  A detailed analysis established a lower
limit on the branching fraction for $Y(4260)\rt \pipi\jp$
of 0.6\%.\cite{Mo_y4260}
This coupled with the measured resonance width
implies that the $Y(4260)$ has a larger partial decay for
$\pipi\jp$ that is at least an order-of-magnitude larger
than those of the  established charmonium states.
Another problem with a charmonium assignment for the $Y(4260)$
is the lack of availability of any unoccupied $1^{--}$ states.
The three $1^{--}$ charmonium states in this mass range,
the $3^3S_1$, $2^3D_1$ and $4^3S_1$ slots have already
been assigned to the $\psi(4040)$, $\psi(4160)$ and
$\psi(4415)$.  The properties of these states, including 
their decay rates to hadrons, match well to charmonium model 
predictions\cite{barnes:2005} and there does not seem to be any
compelling reasons to alter these assignments.

\paragraph{$X(4325)$~~~~}
The $Y(4325)$ was discovered by BaBar as a 
broad enhancement peaking near 4325~MeV  
in the $\pipi\psip$ mass distribution
for $e^+e^-\rt \gamma_{ISR}\pipi\psip$
radiative return events in a 298~fb$^{-1}$
data sample.  This peak
can neither be fitted with a $Y(4260)$ line-shape 
nor does any evidence for it appear in either
the $\pipi\jp$ channel or in the total
cross section measurements.  Since this
also has to have $J^{PC}=1^{--}$, there
are no unfilled charmonium states that it
could be assigned to.

\section{What's new?}

\subsection{$X(3872)$ news}
This past summer.
the Belle group reported the first observation of $X(3872)$
production in neutral $B$ meson decays.\cite{belle_x3872_neutral}
(BaBar had previously reported a $\sim3\sigma$ 
signal.\cite{babar_x3872_neutral})  The Belle 
$X(3872)\rt\pipi\jp$ signals from
charged and neutral $B$ decays  are shown in 
Fig.~\ref{fig:x3872_neutral}.

\includegraphics[width=75mm]{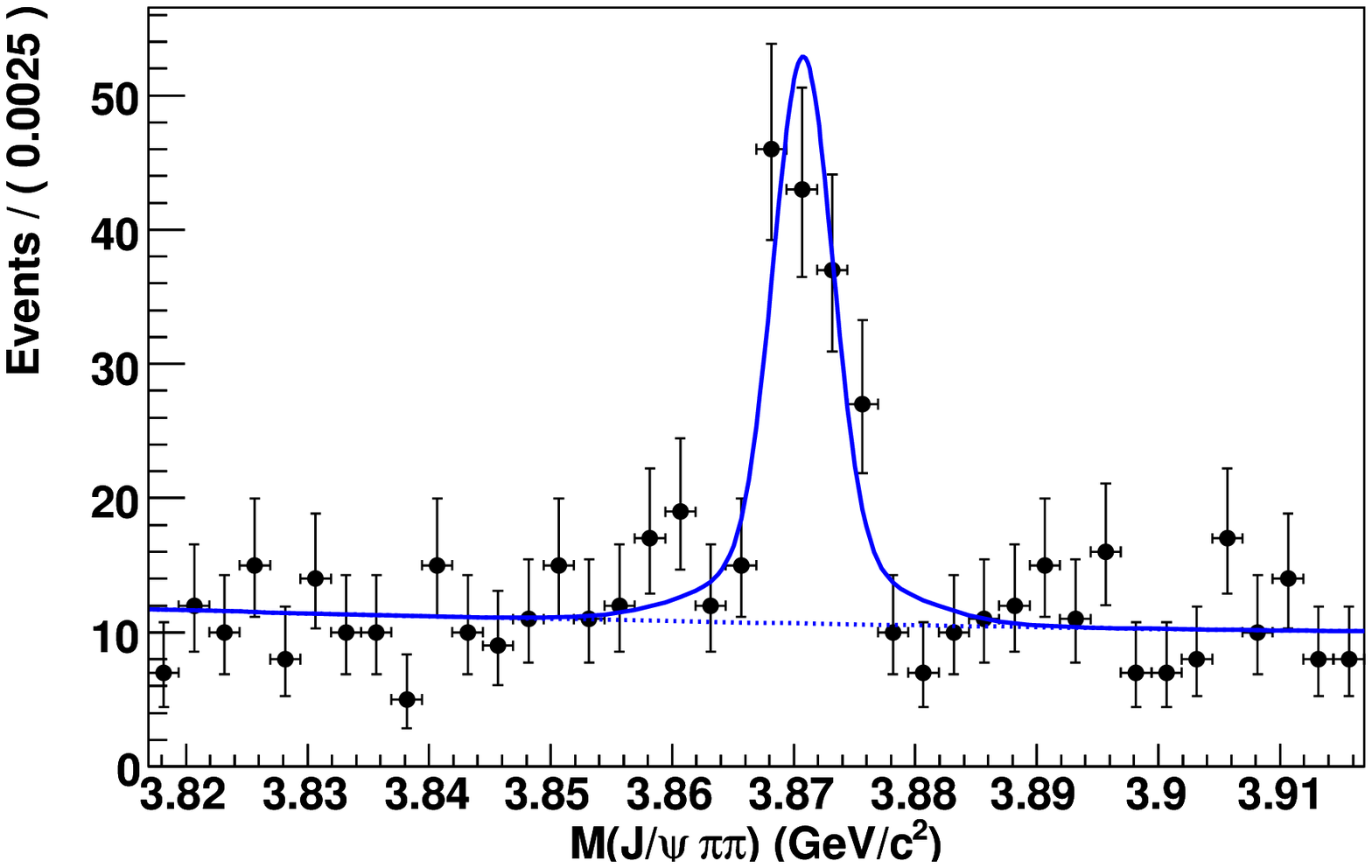}

\includegraphics[width=75mm]{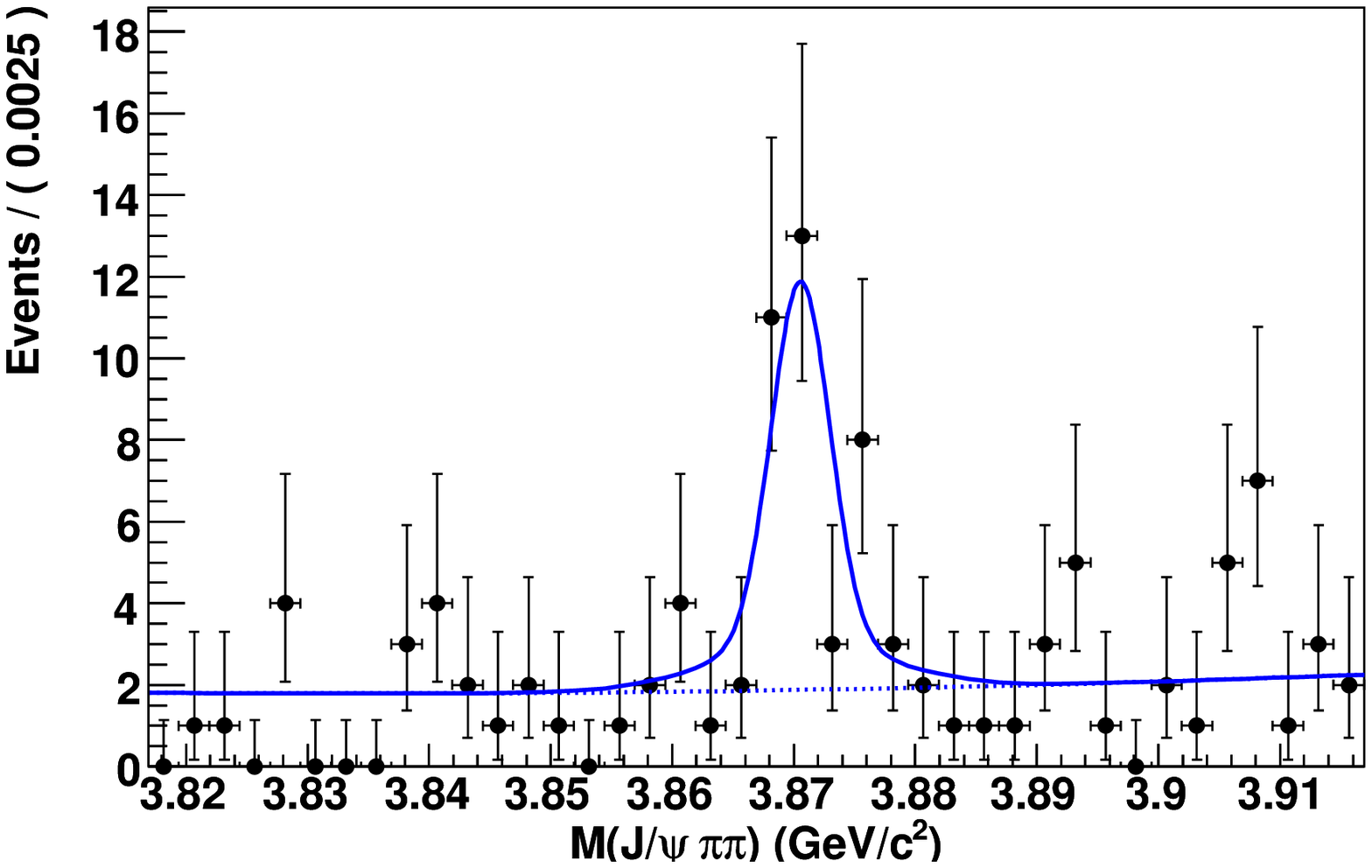}
\figcaption{\label{fig:x3872_neutral} The Belle group's
$X(3872)\rt \pipi\jp$ signals from
{\bf (top)} $B^+\rt K^+\pipi\jp$ and
{\bf (bottom)} $B^0\rt K_S \pipi\jp$ decays.}

Belle finds consistent
mass values for the $X(3872)$ peaks produced from
neutral and charged $B$ decays. Their measured
mass difference for the two sources:\footnote{Here and elsewhere
in this report, the first reported error is statistical and the
second systematic.}
$$
\Delta M = (0.22\pm 0.90~ \pm 0.27~)~{\rm MeV},
$$
is consistent with zero and disagrees with a prediction of 
$(8\pm3)$~MeV by Maiani \etal , which is
based on a diquark-antidiquark model for the 
$X(3872)$ that has a pair of states with 
distinct masses.\cite{diquark}
Belle also reports a ratio of branching fractions
$$
\frac{{\mathcal B}(B^0\rt K^0 X(3872))}{{\mathcal B}(B^+\rt K^+ X(3872))}
= 0.94\pm 0.24 \pm 0.10.
$$
This is consistent with unity, as one might naively expect
based on isospin symmetry.  However, in the 
$D\bar{D^*}$ molecular interpretation of the 
$X(3872)$, large deviations from unity are possible.\cite{braaten:2007}

A second important $X(3872)$-related result reported last summer
was the BaBar group's confirmation of a narrow, near-threshold
peak in the $D \bar{D^*}$ invariant mass distribution for
$B\rt K D\bar{D^*}$ decays (see
Fig.~\ref{fig:x3872_ddpi}).\cite{babar_x3872_ddpi}  
The mass of the 
observed peak
$3875.1 ^{+0.7}_{-0.5}\pm 0.5$~MeV is $4.5\sigma$ higher than
that seen for $X(3872)\rt \pipi\jp$ decays, which confirms
with more significance an earlier result from 
Belle.\cite{belle_x3872_ddpi}   This difference has been
interpreted as being due to threshold effects,\cite{dunwoodie} 
or as evidence for an $X(3872)$ partner state as predicted
by the diquark-antidiquark model.\cite{maiani:2007}

\includegraphics[width=75mm]{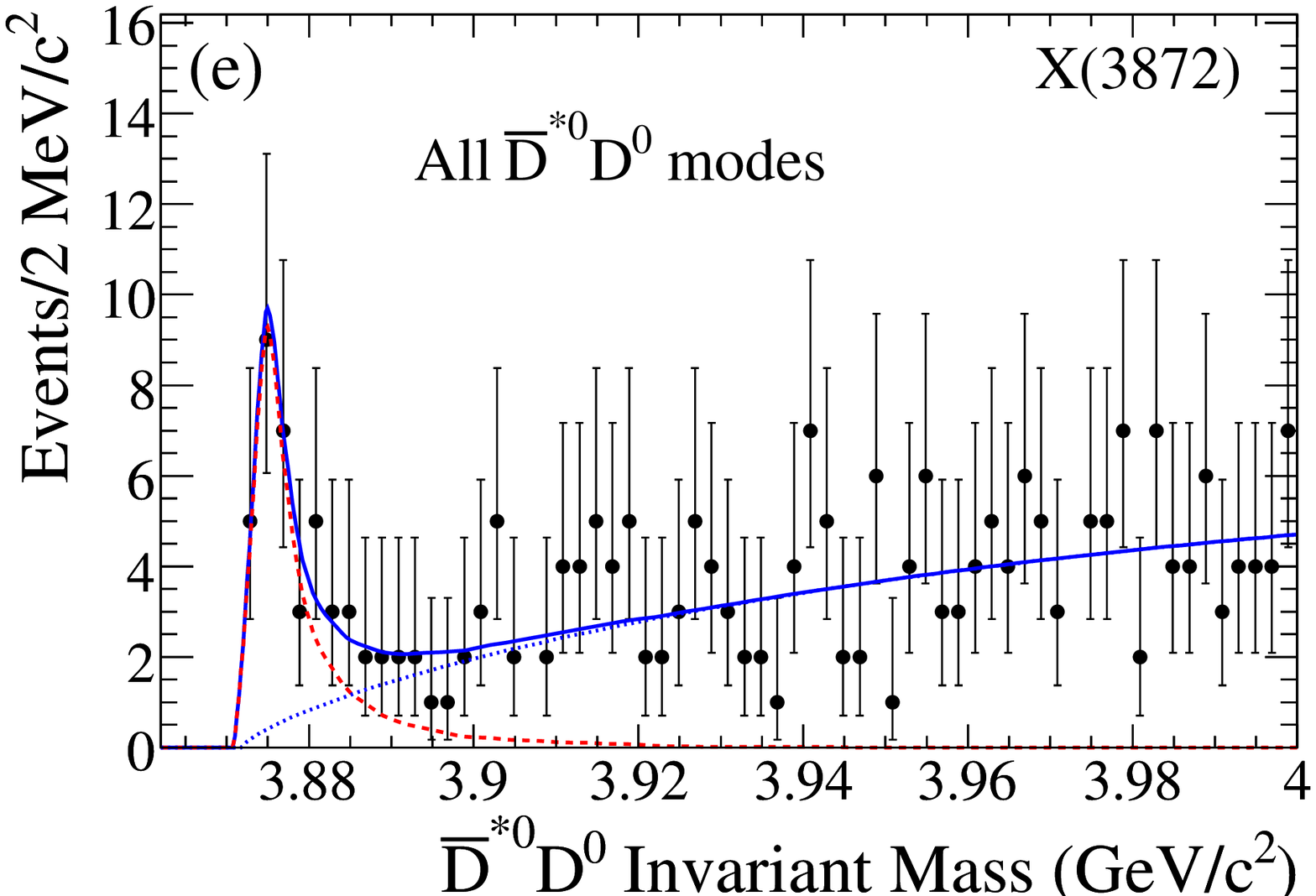}
\figcaption{\label{fig:x3872_ddpi} The
$D \bar{D^*}$ invariant mass distribution for
$B\rt K D\bar{D^*}$ decays (from BaBar).}

\subsection{The $Y(3940)$ is confirmed by BaBar}

This summer the BaBar group reported a study of
$B\rt K\omega\jp$ decays with a data sample containing 383 million 
$B\bar{B}$ meson pairs.\cite{babar_y3940}  
In this sample, the $\omega\jp$ invariant mass spectrum, shown in
Fig.~\ref{fig:babar_y3940}, exhibits a near-threshold enhancement
that is qualitatively similar to the $Y(3940)$ reported earlier 
by Belle.\cite{belle_y3940}   The two groups agree on the product branching 
fraction 
${\mathcal B}(B^+\rt K^+ Y(3940))\times{\mathcal B}(Y(3940)\rt\omega\jp$:
BaBar finds
$(4.9\pm 1.0 \pm 0.5)\times 10^{-5}$ while Belle finds
$(7.1\pm 1.3\pm 3.1)\times 10^{-5}$.  (The Belle result
is an average of the neutral \& charged $B$ samples.)  
However, there is some
disagreement about the mass and width values: BaBar reports
$M=(3914.6 ^{+3.8}_{-3.4} \pm 1.9)$~MeV and $\Gamma = (33 ^{+12}_{-8} 
\pm 5)$~MeV, which are both smaller than Belle's values of
$M=(3943\pm 11\pm 13)$~MeV and $\Gamma = (87 \pm 22 \pm 26)$~MeV.
There are some differences between the two analyses:
BaBar's uses smaller $M(\omega\jp)$ bin sizes
than Belle's: 10~MeV in the  region of the peak versus
40~MeV.  In addition, BaBar exploits the distinctive
Dalitz-plot distribution of $\omega\rt\pipi\pi^0$ decays
by introducing an event-by-event weighting factor. 
However, preliminary results from a reanalysis 
of Belle data with smaller bin sizes and
consideration of the $3\pi$ Dalitz plot distribution
do not resolve the mass and width discrepancies between
the two experiments.  
\vspace{4mm}

\includegraphics[width=60mm]{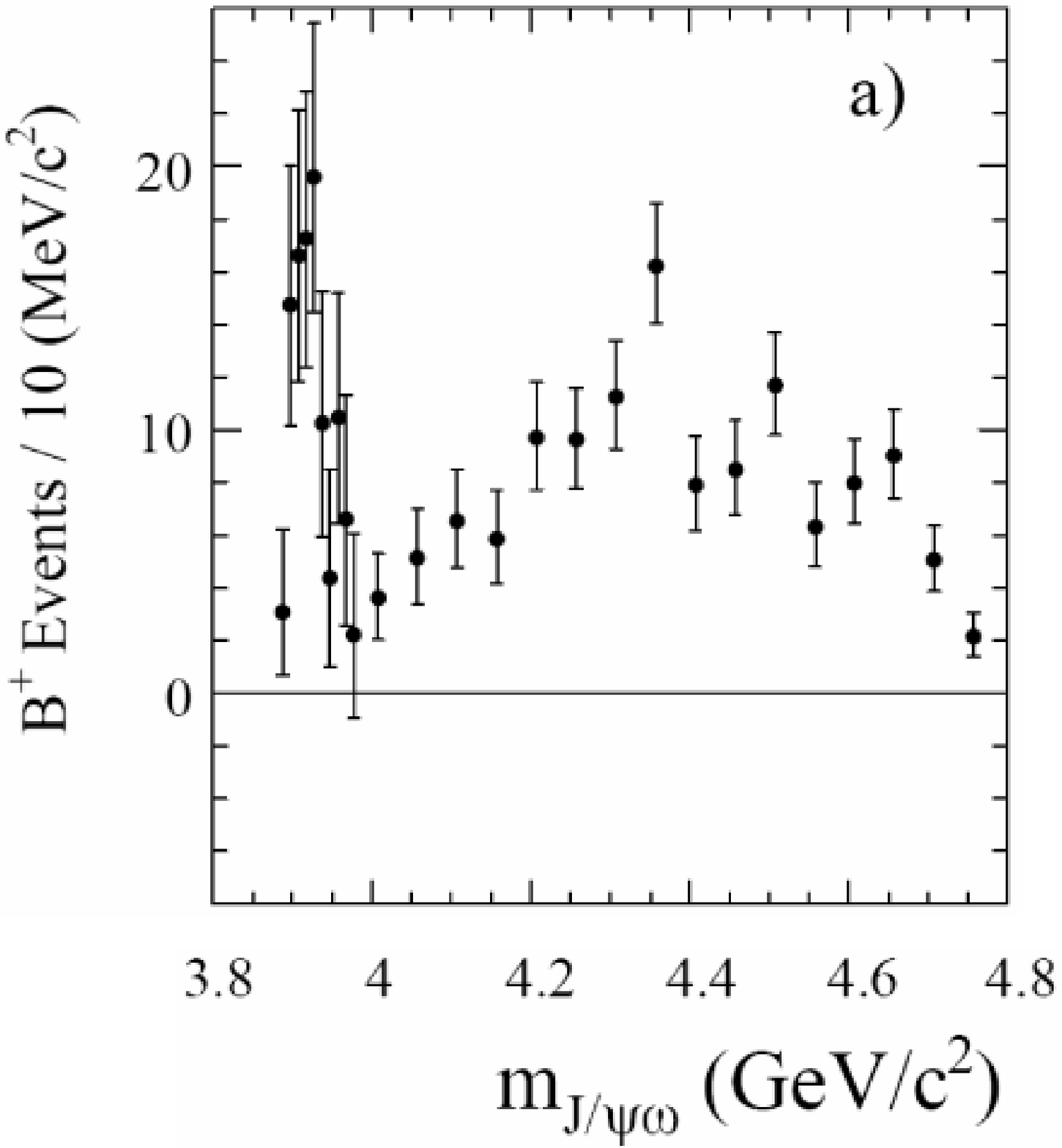}
\figcaption{\label{fig:babar_y3940} The
$\omega\jp$ invariant mass distribution for
$B^+\rt K^+ \omega\jp$ decays (from BaBar).
Below 4.0~GeV, the data are plotted in 10~MeV mass
bins; above 4~GeV, the bin size is 40~MeV.}

\subsection{A new one recoiling from the $\jp$: $X(4160)\rt D^*\bar{D^*}$}

In a continuation of their analyses of $D^{(*)}\bar{D^{(*)}}$
systems recoiling from a $\jp$ in the
$e^+ e^-\rt \jp D^{(*)}\bar{D^{(*)}}$ annihilation process at
$E_{cm} \simeq 10.6$~GeV, Belle confirmed\cite{belle_x4160}
with higher statistics
and better precision their previously reported signal for
$X(3940)\rt D\bar{D^*}$ (see Figs.~\ref{fig:belle_x4160}(b)~and~(c)). 
With a 693~fb$^{-1}$ data sample they
report $M=(3942 ^{+7}_{-6} \pm 6)$~MeV and 
$\Gamma = (37 ^{+26}_{-18} \pm 8)$~MeV.  In addition they report
a $5.5\sigma$ significance signal 
for a new state seen in $D^{*+}D^{*-}$ decays,
(see Fig.~\ref{fig:belle_x4160}(d)) that they
call the $X(4160)$.   The fitted mass and width are
$M=(4156 ^{+25}_{-20} \pm 15)$~MeV and 
$\Gamma = (139 ^{+111}_{-61} \pm 21)$~MeV.  Although the masses
and widths of the $X(4160)$ and the well established 
$\psi(4160)$ $1^{--}$ charmonium state are consistent with each other 
(within errors), they have opposite charge conjugation and, thus, must
be distinct.  Neither $X(3940)$ nor $X(4160)$ signals are evident
in the $D\bar{D}$ invariant
mass distribution for $e^+e^-\rt\jp D\bar{D}$ annihilations,
which is shown in Fig.~\ref{fig:belle_x4160}(a).  Here instead, there is
a broad enhancement above background that peaks around 3880~MeV. 
Although the excess above background is significant, the data are
not sufficient to establish a resonance shape.

As mentioned above, the $X(3940)$ is a candidate for the
$\eta^{\prime\prime}_c$ even though its mass is somewhat
lower than theoretical expectations.  In contrast, the $X(4160)$
mass is higher than $\eta^{\prime\prime}_c$
expectations, and much too low to be the $\eta^{\prime\prime\prime}_c$,
which is expected to have a mass near 4400~MeV.  Thus, 
although it is conceivable that either the $X(3940)$ or the
$X(4160)$ could be a standard $c\bar{c}$ meson,
it seems very unlikely that they both could be 
assigned to charmonium states.

\includegraphics[width=75mm]{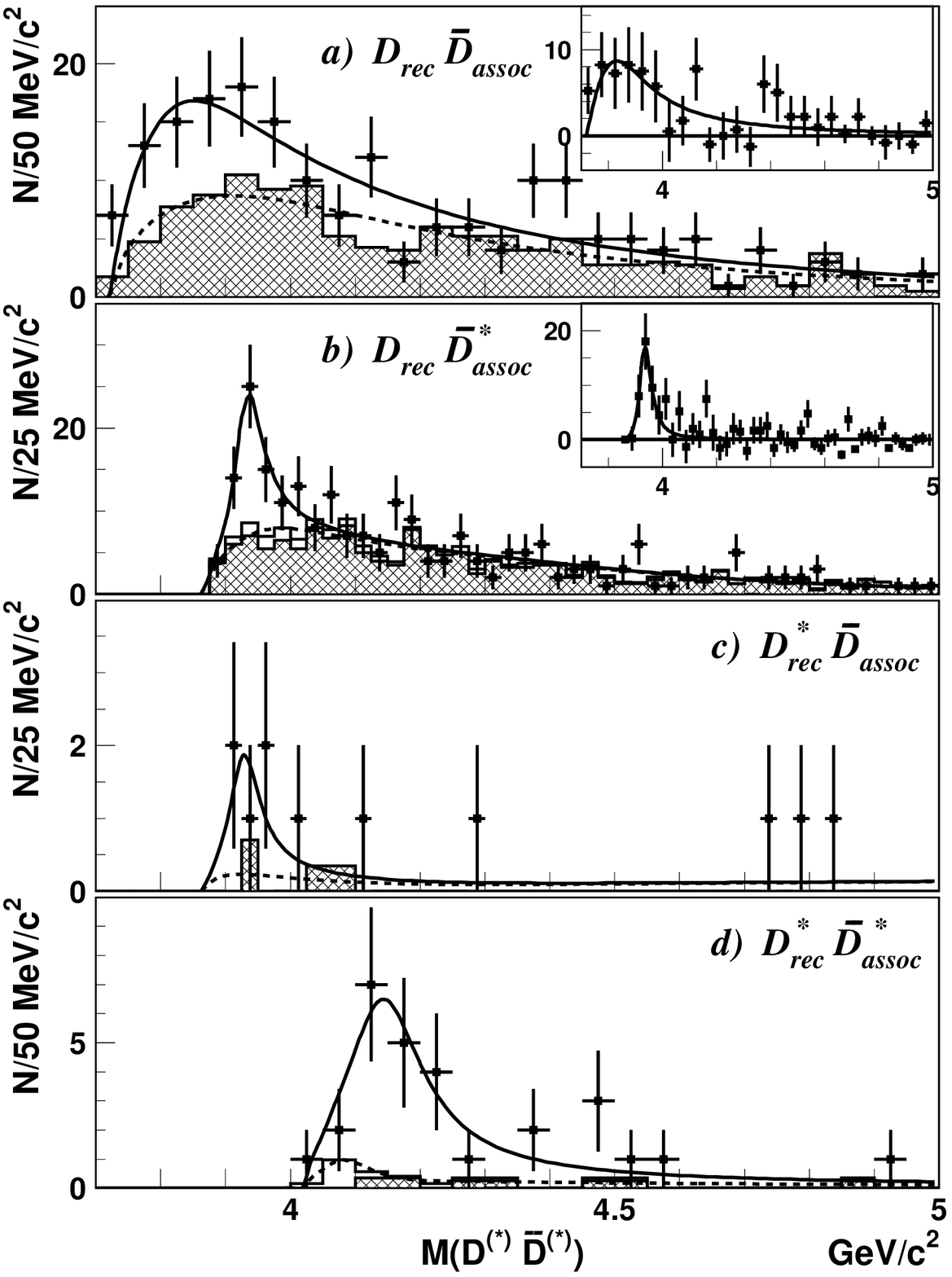}
\figcaption{\label{fig:belle_x4160} 
{\bf (a)} The $D \bar{D}$ invariant mass distribution for
$e^+e^-\rt \jp D\bar{D}$ annihilations. Here
one $D$ is detected and the other is inferred
from kinematics. Panels
{\bf (b)} and {\bf (c)} show the 
$D \bar{D^*}$ invariant mass distribution for
$e^+e^-\rt \jp D\bar{D^*}$. 
In {\bf (b)}, 
the $D$ is detected and the $D^*$ is inferred;
in {\bf (c)}, 
the $D^*$ is detected and the $D$ is inferred.
panel {\bf (d)} shows the 
$D^{*+} D^{*-}$ invariant mass distribution for
$e^+e^-\rt \jp D^{*+} D^{*-}$, where one
$D^*$ is detected and the other inferred. (From Belle.)}

\subsection{News on the $1^{--}$ states}

The top panel of Fig.~\ref{fig:belle_y4260} shows
recently reported Belle results for the 
$\pipi\jp$ invariant mass distribution from a 548~fb$^{-1}$
sample of
$e^+e^-\rt \gamma_{ISR}\pipi\jp$ radiative return 
events.\cite{belle_y4260}
The mass distribution shows a distinct peak with mass and
width of $M=(4247 \pm 12 ^{+17}_{-32})$~MeV and
$\Gamma = (108 \pm 19 \pm 10)$~MeV; results that
confirm BaBar's $Y(4260)$.  In addition, there is an
accumulation of events at lower masses that is significantly higher
than the sideband-determined background level.  A fit of a resonance
shape to this enhancement gives mass and width values of
$M=(4008 \pm 40 ^{+114}_{-28})$~MeV and
$\Gamma = (226 \pm 44 \pm 87)$~MeV.  Although the
mass of this second peak is consistent with that of
the $\psi(4040)$ charmonium state, the fitted width value is
much larger than the world average value for the $\psi(4040)$
($80\pm 10$)~MeV.\cite{PDG}  Currently, it is not clear whether 
or not this is another $XYZ$ state, a threshold effect, or
the $\psi(4040)$.

\includegraphics[width=70mm]{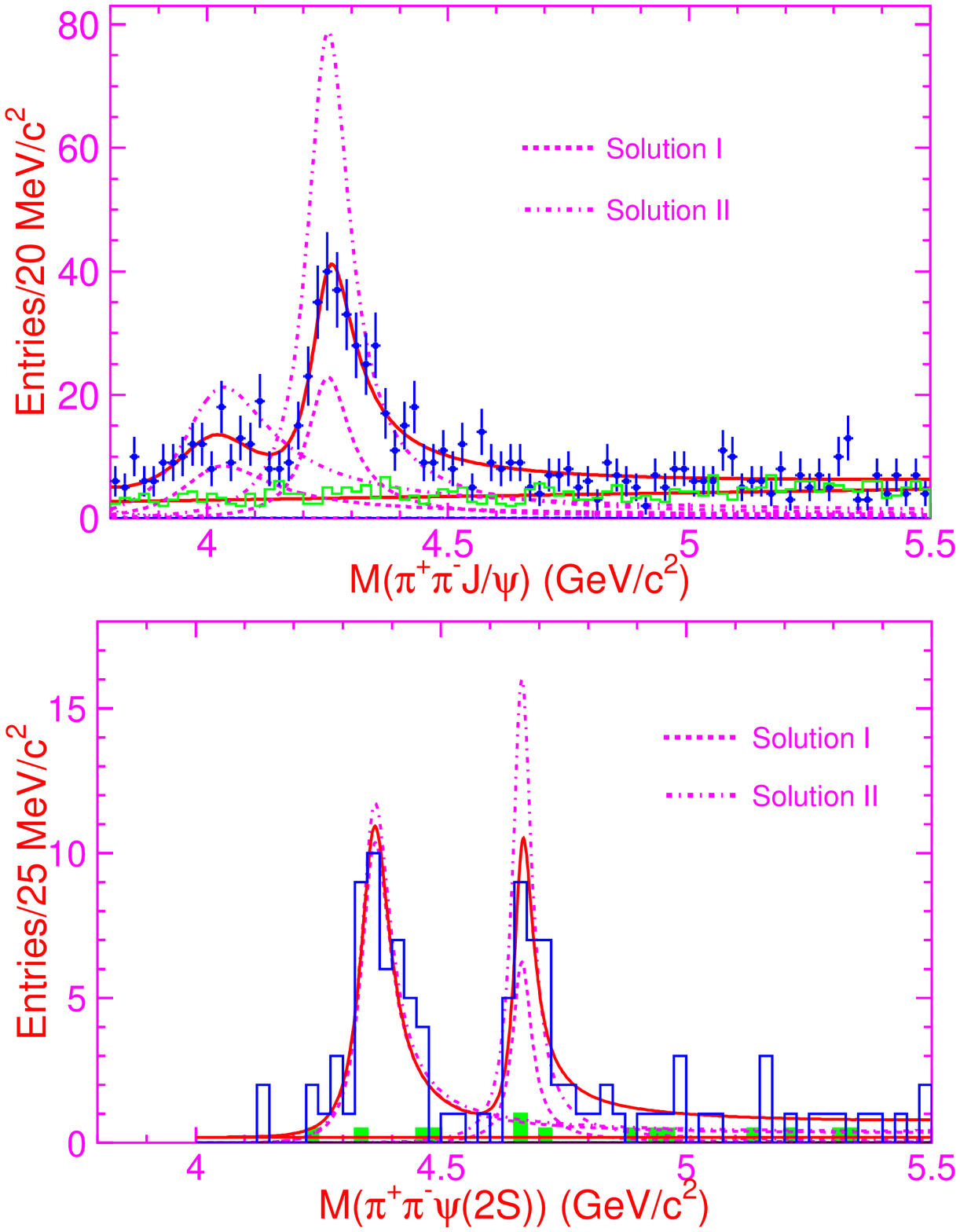}
\figcaption{\label{fig:belle_y4260} The
 $\pipi\jp$ (top) and $\pipi\psip$ (bottom) 
invariant mass distributions
for $e^+e^-\rt \gamma_{ISR}\pipi\jp$ $(\pipi\psip)$
events in Belle.}  

The bottom panel of Fig.~\ref{fig:belle_y4260} shows
the $\pipi\psip$ invariant mass distribution for the
$e^+e^-\rt \gamma_{ISR}\pipi\psip$  
events in a 617~fb$^{-1}$ data sample.\cite{belle_y4660}
Although the data points are consistent with those measured
by BaBar,\cite{babar_y4325}
Belle's larger data sample allows
them to distinguish  two distinct  enhancements:
one at $M= (4361 \pm 9 \pm 9)$~MeV \& 
$\Gamma = (74 \pm 15 \pm 10)$~MeV with a significance
of more than $8\sigma$, and another at  
$M= (4664 \pm 11 \pm 5)$~MeV \& 
$\Gamma = (48 \pm 15 \pm 3)$~MeV with a significance
of $5.8\sigma$.  
Neither peak has parameters consistent with those
of the $Y(4260)$ nor is there any evident signal
for them in the $\pipi\jp$ channel. In addition, there are
no signs of peaking behavior at these masses in 
either the total cross section for $e^+e^-\rt$~hadrons\cite{bes_2003R}
or in the exclusive cross sections for $e^+e^-\rt D\bar{D}$,
$D\bar{D^*}$, $D^*\bar{D^*}$ or $D\bar{D}\pi$.\cite{belle_galina} 
This indicates that the $\pipi\psip$ partial widths are at the $\sim$MeV
level and much larger than those measured for established charmonium 
states ({\it i.e.} $\Gamma(\psip\rt\pipi\jp)=(106 \pm 
4)$~keV and 
$\Gamma(\psi(3770)\rt\pipi\jp) = (49 \pm 8)$~keV\cite{PDG}).

Since all of the $1^{--}$ charmonium states have already been 
assigned, 
the three (or four, if the 4008~MeV peak
is included) $1^{--}$ structures have no available charmonium
assignment.  It has been suggested that coupled-channel effects and 
rescattering between pairs of charmed mesons may be playing a 
role.\cite{voloshin_rescattering}  
However, the peak masses do not overlap
with any of the $D^{(*)}\bar{D^{(*)}}$ or $D_s^{(*)}\bar{D_s^{(*)}}$
thresholds (see Fig.~\ref{fig:thresh}).

A popular interpretation for these $1^{--}$ states is that they
are $c\bar{c}$-gluon hybrids.\cite{mlyan_y4660}  
Lattice-QCD says that the lowest 
mass should be about 4.2~GeV.  In addition, decays to 
$D^{(*)}\bar{D^{(*)}}$ are expected to be suppressed and the
relevant open charm threshold is $m_D + m_{D_1^{**}}\simeq 4285$~MeV, 
which is higher than the peak mass of the $Y(4260)$.  However,
both the $Y(4360)$ and the $Y(4660)$ are well above this
threshold, as is a large part of the high mass tail of the 
$Y(4260)$.  Thus,
the absence of any sign of these states in the total cross
section for $e^+e^-\rt$~hadrons is a problem for the hybrid
interpretation.

\includegraphics[width=75mm]{thresh.epsi}
\figcaption{\label{fig:thresh} The
relation between the various $XYZ$ masses and
charmed-anticharmed meson 
mass thresholds.}

\section{The $Z^+(4430)$}

All of the original $XYZ$ meson candidates were
electrically neutral.  This changed in Summer 2007, 
when Belle reported the observation of a distinct
peak in the $\pi^+\psip$ mass distribution 
produced in $B\rt K\pi^+\psip$ decays (see Fig.~\ref{fig:belle_z4430}).
A fit to this distribution
with an $S$-wave Breit Wigner resonance line-shape
gives resonance parameters of
$M= (4433 \pm 4 \pm 2)$~MeV \& 
$\Gamma = (45 ^{+18} _{-13}{\rm (stat)} ^{+30} _{-13}{\rm (syst)} )$~MeV 
with a signal significance of $6.5\sigma$.\cite{belle_z4430}

In the three-body decay $B\rt K\pi\psip$, the $M(\pi\psip)$ values
are strongly correlated with $\cos\theta_{\pi}$, where
$\theta_{\pi}$ is the angle between the $\pi$ meson and the
$\psip$ directions in the $K\pi$ restframe.  Thus, interference
between different partial waves in the $K\pi$ system that
produce peaks in $\cos\theta_{\pi}$ will produce corresponding
structures in the $M(\pi\psip)$ distribution.  From the
$M(K\pi)$ distribution for these decays, strong contributions
from $S$-wave and $P$-wave $K\pi$ partial waves are evident.
In addition, there is some evidence for the $D$-wave 
$K_2^*(1430)$ resonance state.  However, 
the fitted peak mass value of 4433~MeV corresponds to 
$\cos\theta_{\pi} \simeq 0.25$, and it is not possible to
produce a peak at this value of $\cos\theta_{\pi}$ with
only $S$- $P$- and $D$-waves, without producing additional,
more dramatic structures at other $M(\pi\psip)$ values.
Since no such additional structures are evident in the 
$M(\pi\psip)$ distribution
shown in Fig.~\ref{fig:belle_z4430}, Belle concludes that
the peak they observe is inherent to the $\pi\psip$
system and not a reflection of 
interference effects between different $K\pi$ partial waves.  

\includegraphics[width=75mm]{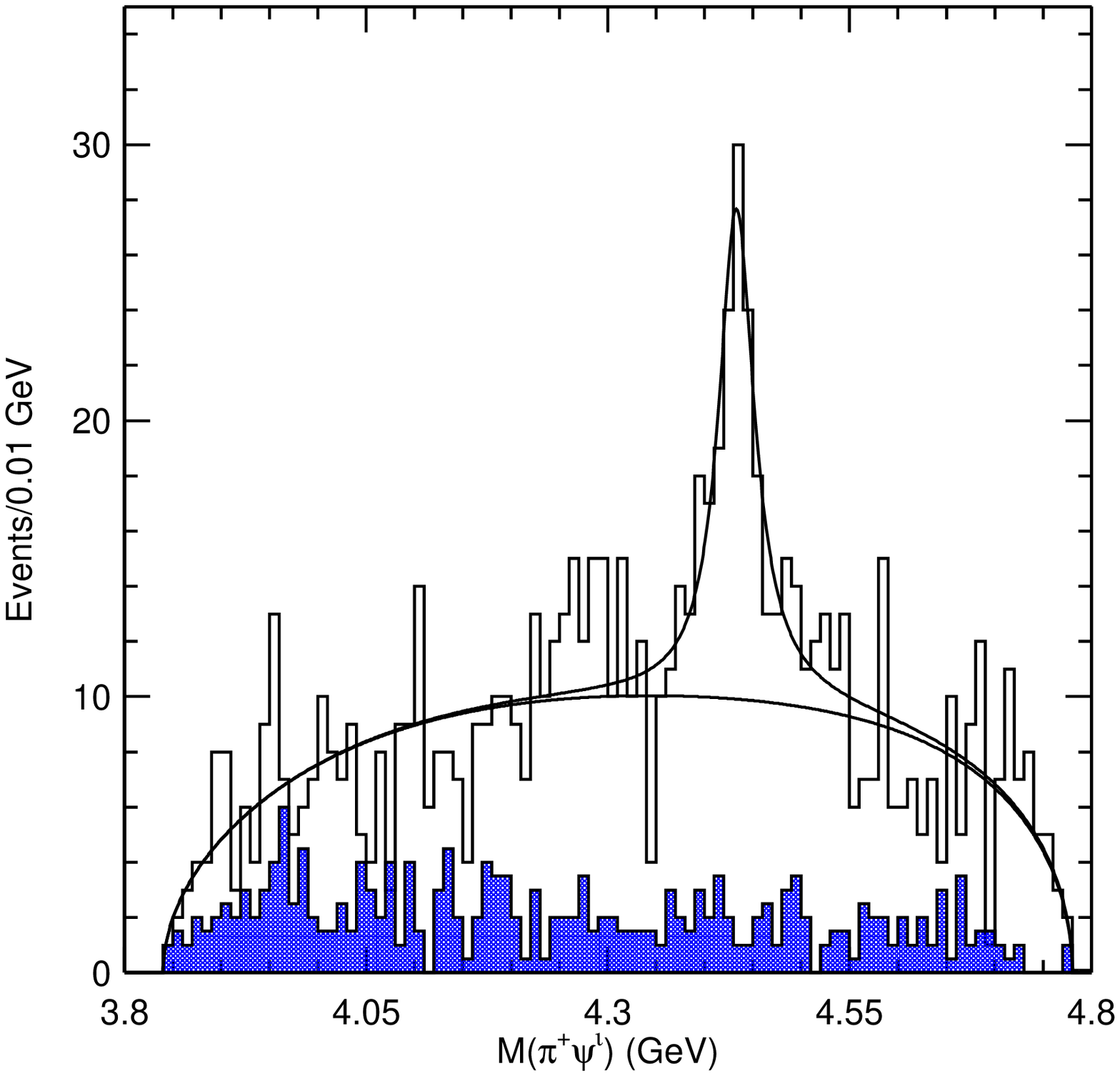}
\figcaption{\label{fig:belle_z4430} The
$M(\pi^{\pm}\psip)$ distribution for $B\rt K\pi^{\pm}\psip$ 
decays (from Belle.)}

If the $Z(4430)$ peak is interpreted as a meson, it
cannot be a $c\bar{c}$ charmonium or a $c\bar{c}$-gluon hybrid
because these are necessarily electrically neutral.  The
remaining possibility is a tetraquark state.  
Some authors have proposed
that it is a $D^*\bar{D_1^{**}}$ molecule\cite{molecule_z4430}
and others have advocated a diquark-antidiquark 
interpretation ({\it e.g.} 
a $(cu)(\bar{c}\bar{d})$ combination).\cite{maiani:2007a}
The $Z(4430)$ bears some similarities to the $Y(4360)$
and $Y(4660)$ in that they are in the same mass range,
have similar widths and are observed to decay to $\psip$
and not $\jp$.  If, in fact, they are related, this would
cause trouble with the hybrid interpretation for the
$Y(4360)$ and $Y(4660)$, as well as with the  
$D^*\bar{D_1^{**}}$ molecule
picture for the $Z(4430)$.

\section{Are the corresponding states in the $s$- and $b$-quark sectors?}

The proliferation of meson candidates that are strongly coupled
to $c\bar{c}$ quark pairs but not compatible with a conventional
charmonium assignment leads one naturally to question whether or
not similar states exist that are strongly coupled to $s\bar{s}$
or $b\bar{b}$ quark pairs. There is some evidence that this, in fact,
may be the case. 

\subsection{The $Y(2175)$}
In 2006, the BaBar group reported a resonance-like structure in the
$f_0(980)\phi$ invariant mass distribution produced in
$e^+e^-\rt \gamma_{ISR} f_0(980)\phi$ radiative-return 
events.\cite{babar_y2175}  They report resonance parameters of
$M= (2170 \pm 10 \pm 15)$~MeV \& 
$\Gamma = (58 \pm 16  \pm 20 )$~MeV. They see no signal for
this peak in a sample of $K^*(892)K\pi$
events that has little kinematic overlap with $f_0(980)\phi$, and
conclude that this structure, which they call the $Y(2175)$,
has a relatively large branching fraction for $f_0(980)\phi$.

\includegraphics[width=65mm]{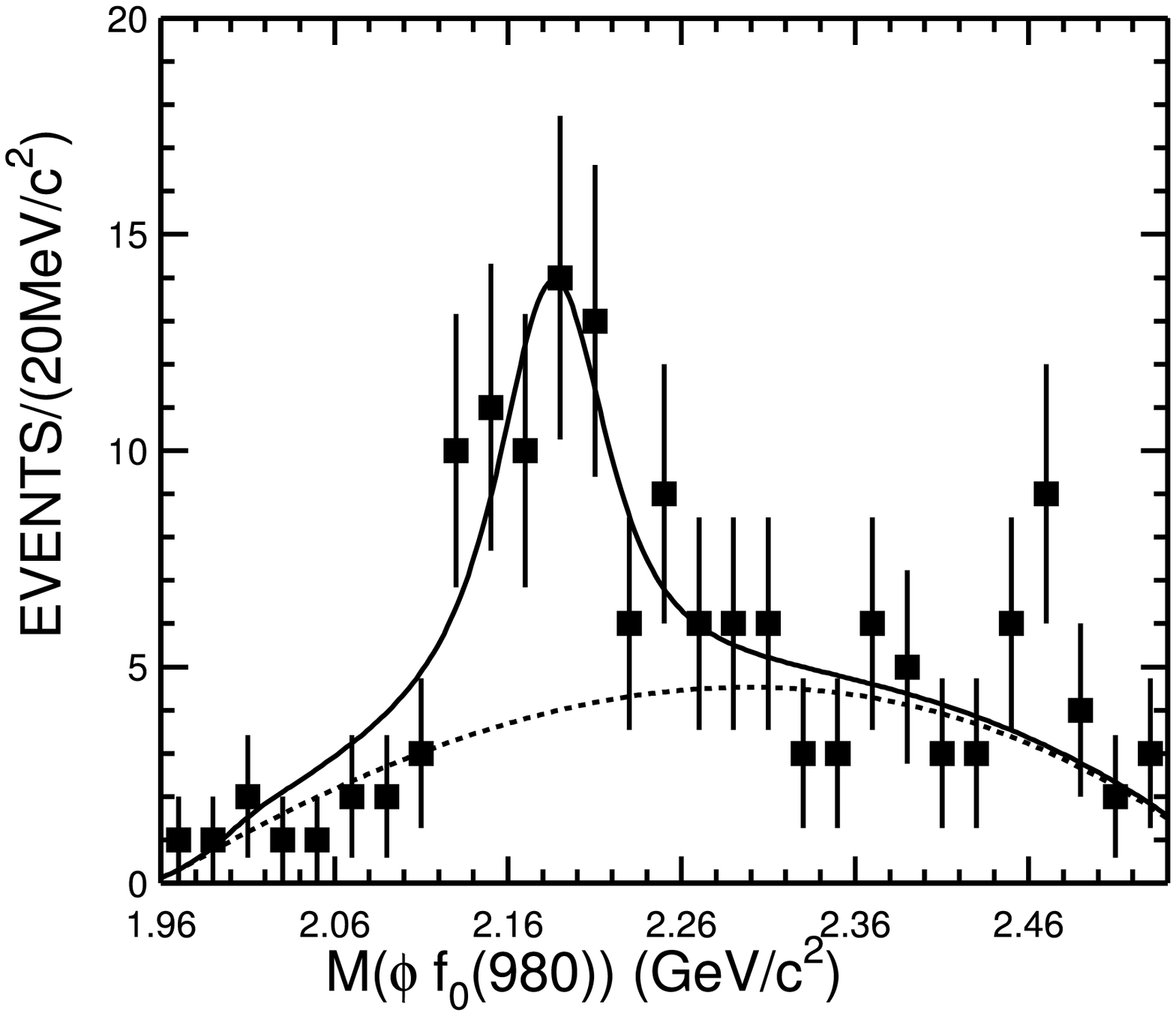}
\figcaption{\label{fig:bes_y2175} The
$M(f_0(980)\phi)$ distribution for $\jp\rt \eta f_0(980)\phi$ decays
in BESII.}

The similarities with the $Y(4260)$, both in production and
decay properties,
led them to speculate that the $Y(2175)$ might be an
$s\bar{s}$ analogue of the $Y(4260)$, 
{\it i.e.} it is the ``$Y_s(2175)$''.  
On the other hand, 
there is no compelling evidence against it being a
conventional $3^3S_1$ or  $2^3D_1$ $s\bar{s}$ 
``strangeonium''  state.  The study of the $Y(2175)$ in 
other production
and decay modes would be useful for distinguishing 
between different possibilities.\cite{ding_y2175}

The BESII group made a first step in this program 
by finding an $f_0(980)\phi$  mass peak with 
similar parameters produced in 
$\jp\rt\eta f_0(980)\phi$ decays 
(see Fig.~\ref{fig:bes_y2175}).\cite{bes_y2175}  
The BESII fit yields a mass and width of
$M= (2186 \pm 10 \pm 6)$~MeV \& 
$\Gamma = (65 \pm 23  \pm 17 )$~MeV, which are in
good agreement with BaBar's measurements.

The next
steps will be finding it in other decay modes and
searching for counterpart states  with quantum numbers
other than $1^{--}$  that, perhaps, decay into final
states containing an $\eta^{\prime}$.  This will be
an important task for BESIII.

\subsection{Anomalous $\pipi\Upsilon(nS)$ production at the 
$\Upsilon(5S)$}

Using a sample of 236 million $\Upsilon(4S)$ mesons, 
BaBar\cite{babar_4s2pipi1s} observed $167 \pm 19$
and $97 \pm 15$ event signals for
$\Upsilon(4S)\rt\pipi\Upsilon(1S)$ and $\pipi\Upsilon(2S)$, 
respectively, from which they infer partial widths
$\Gamma(\Upsilon(4S)\rt\pipi\Upsilon(1S)) = (1.8 \pm 0.4)$~keV.
$\Gamma(\Upsilon(4S)\rt\pipi\Upsilon(2S)) = (2.7 \pm 0.8)$~keV.
Belle\cite{belle_4s2pipi1s}, with a sample of 464 million $\Upsilon(4S)$
events  reported a $44 \pm 8$ event signal for the transition 
$\Upsilon(4S)\rt\pipi\Upsilon(1S)$, from which they infer a partial
width 
$\Gamma(\Upsilon(4S)\rt\pipi\Upsilon(1S)) = (3.65 \pm 0.67 \pm 0.65)$~keV.
These partial widths are comparable in magnitude to those measured
for $\pipi$ transitions between the $\Upsilon(3S)$, $\Upsilon(2S)$
and $\Upsilon(1S)$.\cite{PDG}

In 2006, Belle had a one-month-long run at $e^+e^-$ cm energy of 
10.87~GeV,
which corresponds to the peak mass of the $\Upsilon(5S)$.  The total
data sample collected was 21.7~fb$^{-1}$ and the
number of $\Upsilon(5S)$ events collected was 6.3 million.
Much to their surprise, they found large numbers of $\pipi\Upsilon(nS)$
events in this data sample: $325 \pm 20$ $\pipi\Upsilon(1S)$ events and
 $186 \pm 15$ $\pipi\Upsilon(2S)$ events (see 
Figs.~\ref{fig:belle_5s2pipi1s_fig2a}(a) and (b)).\cite{belle_5s2pipi1s}
(The $\Upsilon(2,3S)\rt\pipi\Upsilon(1S)$ signals 
in Fig.~\ref{fig:belle_5s2pipi1s_fig2a}(a) are
produced by radiative-return transitions 
$e^+e^-\gamma_{ISR}\Upsilon(2,3S)$.)

\includegraphics[width=75mm]{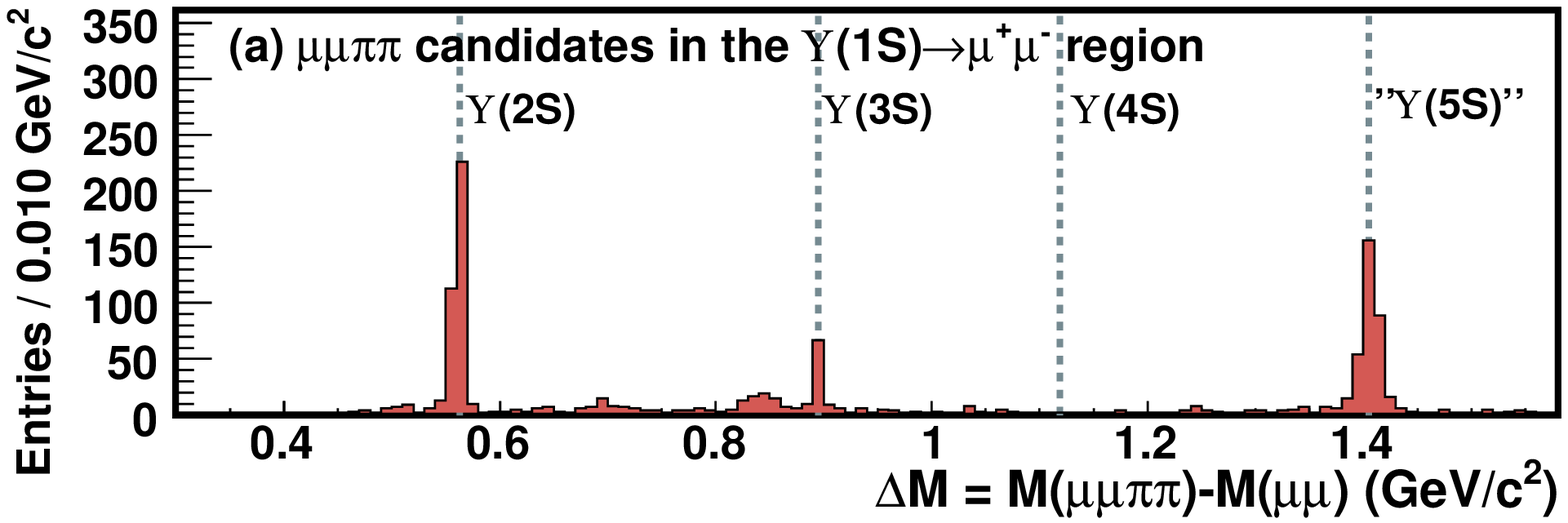}

\includegraphics[width=75mm]{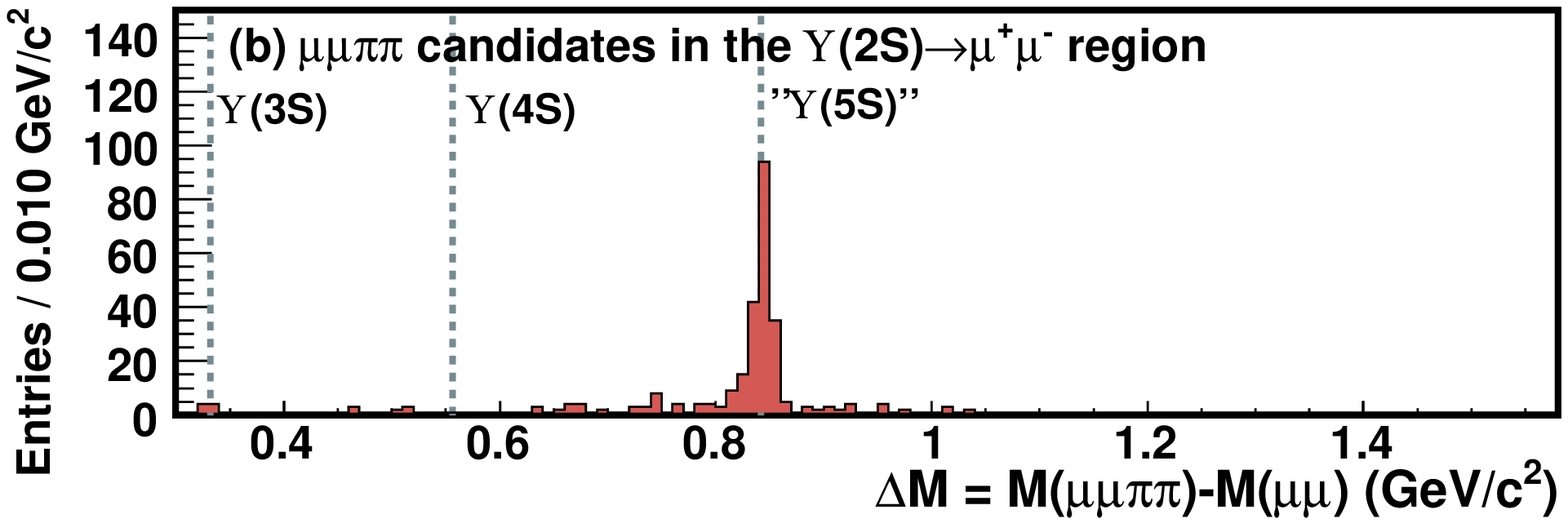}

\figcaption{\label{fig:belle_5s2pipi1s_fig2a} 
Belle's
$M(\mumu\pipi)-M(\mumu)$ mass difference distributions
for events with {\bf (a)} $M(\mumu)=\Upsilon(1S)$
and {\bf (b)} $M(\mumu)=\Upsilon(2S)$. Vertical dashed
lines show the expected locations for 
$\Upsilon(nS)\rt \pipi\Upsilon(1,2S)$ transitions.}

If one assumes that these events
are coming from $\Upsilon(5S)\rt\pipi\Upsilon(nS)$ transitions, the
inferred partial widths are huge:
$\Gamma(\Upsilon(5S)\rt\pipi\Upsilon(1S)) = (590 \pm 40 \pm 90)$~keV.
$\Gamma(\Upsilon(5S)\rt\pipi\Upsilon(2S)) = (850 \pm 70 \pm 160)$~keV,
more than two orders-of-magnitude higher than corresponding transitions
from the $\Upsilon(4S)$.

A likely explanation for these unexpectedly large partial
widths (and, in fact, the motivation for
Belle's pursuit of this subject) is that there 
is a ``$Y_b$'', {\it i.e.} a $b\bar{b}$ counterpart of the
$Y(4260)$, that is overlapping the $\Upsilon(5S)$,\cite{hou_yb} 
and this
state is producing the $\pipi\Upsilon(1,2S)$ events that are seen.
To test this possibility, Belle performed an energy scan to
map the $\pipi\Upsilon(1,2S)$ cross section in the cm energy region
around $10.87$~GeV during December 2007.  Results from 
this scan will be reported in the near future.

\section{Summary}

A large (and growing) number of candidate charmonium-like
meson states have been observed
that do not seem to fit into the quark-antiquark classification scheme
of the constituent quark model.  Some of the salient properties of the
states discussed in this report are summarized in Table~\ref{table1},
which is modeled after the one shown by Eichten at QWG2007.  

These states exhibit a  number of peculiar features:
\begin{itemize}
\item
Many of them have partial widths for decays to charmonium~+~light~hadrons
that are at the $\sim$MeV scale, which is much larger than is typical
for established $c\bar{c}$ meson states.
\item
They are relatively narrow although many of them are well above
relevant open-charm thresholds.
\item
There seems to be some selectivity: states seen to decay to final
states with a $\psip$ are not seen in the corresponding $\jp$
channel, and {\it vice versa}.
\item
The new $1^{--}$ charmonium states are not apparent in the 
$e^+e^-\rt$ charmed-meson-pair or the total hadronic cross sections.
\item 
There are no evident changes in the properties of these states 
at the $D^*D^{**}$ mass threshold.
\item
Although some states are near mass thresholds for pairs of open charmed
mesons, this is not a universal feature (see Fig.~\ref{fig:thresh}).
\item
There is some evidence that similar states exist in the $s$- and $b$-quark
sectors.
\end{itemize}

Attempts to explain these states theoretically have usually been
confined to subsets of the observed states.  For example,
the $X(3872)$ and $Z(4430)$ have been attributed to bound 
molecular states of $D\bar{D^*}$ and $D^*\bar{D^{**}}$ mesons, 
or as  diquark-antidiquark tetraquark states,
the $Y(4260)$ as a $c\bar{c}$-gluon hybrid, etc.
However, no single model seems able to deal with the whole system 
and their properties in a compelling way.  In general, the predictions 
of the various models have had limited success.

This continues to be a data-driven field, with an increasingly 
large number of new results continuing to come out from BaBar, Belle and BES.
Hopefully, this deluge of information will eventually lead to
a more unified and clearer picture of what is going on. 

\end{multicols}
\begin{center}
\tabcaption{\label{table1} Summary of the candidate $XYZ$ mesons discussed 
in the text.}
\footnotesize
\begin{tabular*}{180mm}{lcccccc}
\hline\hline
state    & $M$~(MeV) &$\Gamma$~(MeV)    & $J^{PC}$ & Decay Modes        & 
Production  Modes & Observed by:\\\hline
$Y_s(2175)$& $2175\pm8$&$ 58\pm26 $& $1^{--}$ & $\phi f_0(980)$     &  
$\ee$~(ISR), $\jp\rt\eta Y_s(2175)$  &  BaBar, BESII\\
$X(3872)$& $3871.4\pm0.6$&$<2.3$& $1^{++}$ & $\pipi \jp$,$\gamma \jp$
 & $B\rt KX(3872)$, $p\bar{p}$ & Belle, CDF, D0, BaBar\\
$X(3875)$& $3875.5\pm 1.5$&$3.0^{+2.1}_{-1.7}$  & ?  &$D^0\bar{D^0}\pi^0 (\gamma)$ 
& $B\rt K X(3875)$ &  Belle, BaBar \\
$Z(3940)$& $3929\pm5$&$ 29\pm10 $& $2^{++}$ & $D\bar{D}$   & $\gamma\gamma\rt Z(3940)$ 
& Belle \\
$X(3940)$& $3942\pm9$&$ 37\pm17 $& $J^{P+}$ & $D\bar{D^*}$ & $\ee\rt \jp X(3940)$
& Belle  \\
$Y(3940)$& $3943\pm17$&$ 87\pm34 $&$J^{P+}$ & $\omega J/\psi$  & $B\rt K Y(3940)$ 
& Belle, BaBar \\
$Y(4008)$& $4008^{+82}_{-49}$&$ 226^{+97}_{-80}$ &$1^{--}$& $\pipi \jp$ & $\ee$(ISR) 
& Belle \\
$X(4160)$& $4156\pm29$&$ 139^{+113}_{-65}$ &$J^{P+}$& $D^*\bar{D^*}$& 
$\ee \rt \jp X(4160)$  & Belle \\ 
$Y(4260)$& $4264\pm12$&$ 83\pm22$ &$1^{--}$&  $\pipi \jp$ & $\ee$(ISR) 
&BaBar, CLEO, Belle    \\
$Y(4350)$& $4361\pm13$&$ 74\pm18$ &$1^{--}$&  $\pipi \psip$ & $\ee$(ISR) 
& BaBar, Belle  \\
$Z(4430)$& $4433\pm5$&$ 45^{+35}_{-18}$ & ? & $\pi^{\pm}\psip$ & $B\rt K Z^{\pm}(4430)$
& Belle \\
$Y(4660)$& $4664\pm12$&$ 48\pm15 $ &$1^{--}$&  $\pipi \psip$ & $\ee $(ISR) 
& Belle     \\
$Y_b$  & $\sim 10,870$ & ? &   $1^{--}$ & $\pipi\Upsilon(1,2S)$ 
& $\ee\rt Y_b$ & Belle \\
\hline\hline
\end{tabular*}%
\end{center}


\begin{multicols}{2}

\section{Acknowledgements}

I thank Changchun Zhang and Ying Hua Jia for organizing
this interesting and useful meeting.
I was assisted in preparing this report by Sookyung Choi of Gyeongsang
National University,
Changzheng Yuan and Xia Wang of
IHEP-Beijing, Kai-Feng Chen of National Taiwan University and Karim 
Trabelsi 
of KEK.  This work is supported by the
Experimental Physics Center of IHEP-Beijing and the U.S. Department of Energy.

\end{multicols}


\vspace{5mm}

\begin{thebibliography}{90}


\bibitem{skchoi_etac2s} Choi S-K {\etal} (Belle Collaboration).  
Phys. Rev. Lett. )(2002) {\bf 89}:102001.

\bibitem{belle_z3940}Uehara~S {\etal} (Belle Collaboration).
Phys. Rev. Lett. (2005) {\bf 96}:082003.

\bibitem{belle_x3872} Choi~S-K {\etal} (Belle Collaboration).  
Phys. Rev. Lett. (2003) {\bf 91}:262001.

\bibitem{babar_y4260} Aubert~B {\etal} (BaBar Collaboration).  
Phys. Rev. Lett. (2005) {\bf 95}:142001.

\bibitem{belle_x3940} Abe~K {\etal} (Belle Collaboration).
Phys. Rev. Lett. (2007) {\bf 98}:082001.

\bibitem{belle_y3940} Choi~S-K {\etal} (Belle Collaboration).  
Phys. Rev. Lett. (2005) {\bf 94}:182002.

\bibitem{babar_y4325} Aubert~B {\etal} (BaBar Collaboration).
Phys. Rev. Lett. (2007) {\bf 98}:212001.


\bibitem{molecular}See, for example, 
Voloshin M B, Okun L B.
JETP Lett. (1976) {\bf 23}:333;
Bander M, Shaw G L, Thomas P.
Phys. Rev. Lett. (1977) {\bf 36}:695;
De~Rujula A, Georgi H, Glashow S L.  
Phys. Rev. Lett. (1977) {\bf 38}:317; 
T\"{o}rnqvist N A. Z. Phys. C (1994) {\bf 61}:525; 
Manohar A V, Wise M B. Nucl. Phys. B (1993) {\bf 339}:17;
T\"{o}rnqvist N A., Phys. Lett.(2004) {\bf B590}:209; 
Close F E, Page P R. Phys. Lett. B (2003) {\bf 578}:119;
Wong C Y. Phys. Rev. C (2004) {\bf 69}:055202; 
Braaten E, Kusunoki M. Phys. Rev. D (2004) {\bf 69}:114012;
Swanson E S. Phys. Lett. (2004) {\bf B588}:189;
Gamermann E, Oset E. Eur. Phys. J (2007) {\bf A33}:119-131
and arXiv:0712.1758.

\bibitem{diquark} 
Chiu T-W, Hsieh T-H. Phys. Lett. (2007) {\bf B646}:95;
Bigi I, Maiani L, Piccinini F, Polosa A D, Riquer V.  
Phys. Rev. D (2005) {\bf 72}:114016;
Maiani L, Riquer V, Piccinini F, Polosa A D. 
Phys. Rev. D (2005) {\bf 72}:031502(R).

\bibitem{hybrid} 
Horn D, Mandula J Phys. Rev. D (1978) {\bf 17}:898; 
Close F E. Phys. Lett. B (1995) {\bf 342}:369; 
McNeile C, Michael C, Pennanen P.
Phys. Rev.~D (2002) {\bf 65}:094505.

\bibitem{eichten} 
Barnes T, Godfrey S. Phys. Rev. D (2004) {\bf 69}:054008; 
Eichten E J, Lane K, Quigg C. Phys. Rev. D (2004)
{\bf 69}:094019; Voloshin M B, Phys. Lett. (2004) {\bf B579}:316.

\bibitem{chao} Meng C, Chao K T.
Phys. Rev. D (2007) {\bf 75}:114002.

\bibitem{chiu} Chiu T-W, Hsieh T-H.  
Phys. Rev. D (2006) {\bf 73}:111503(R).


\bibitem{CDF_JPC} Abulencia~D {\etal} (CDF Collaboration).
Phys. Rev. Lett. (2007) {\bf 98}:132002.

\bibitem{Belle_JPC} Abe~K {\etal} (Belle Collaboration).
arXiv:hep-ex/0505038 (2005).

\bibitem{barnes:2005} 
Barnes T, Godfrey S, Swanson E S. Phys. Rev. D (2004) {\bf 72}:054026.

\bibitem{quigg:2006} 
Eichten E J, Lane K, Quigg C. Phys. Rev. D (2006) {\bf 73}:014014. 


\bibitem{BaBar_B2jpsi-incl} Aubert B {\etal} (BaBar Collaboration).
Phys. Rev. Lett. (2006) {\bf 96}:052002.

\bibitem{bes_2003R} Bai J Z {\etal} (BESII Collaboration).
Phys. Rev. Lett. (2002) {\bf 88}:102802.

\bibitem{CB_R} Osterfeld A {\etal} (Crystal Ball Collaboration).
Report No. SLAC-PUB-4160 (1986) (unpublished).

\bibitem{Mo_y4260}
Mo~X H, Li G, Yuan C Z, Hu H M, Hu J H, Wang P, Wang Z Y.
Phys. Lett. B (2006) {\bf 640}:182-189.

\bibitem{belle_x3872_neutral} Abe K {\etal} (Belle Collaboration).
Belle-CONF-0711 August 2007.

\bibitem{babar_x3872_neutral} Aubert B {\etal} (BaBar Collaboration).
Phys. Rev. D (2006) {\bf 73}:011101(R).


\bibitem{braaten:2007}
Braaten E, Lu M. arXiv:0710.5482 [hep-ph].

\bibitem{babar_x3872_ddpi}
 Aubert B {\etal} (BaBar Collaboration).
arXiv:0708.1565 [hep-ex] (2007) submitted to Physical Review D.

\bibitem{belle_x3872_ddpi} Gokhroo G {\etal} (Belle Collaboration). 
Phys. Rev. Lett. (2006) {\bf 97}:162002.

\bibitem{dunwoodie}
Dunwoodie W, Ziegler V. arXiv:0710.5191 (2007) [hep-ex];
Hanhart C, Kalashnikova Yu S, Kudryavtsev A E, Nefediev A V.
Phys. Rev. D (2007) {\bf 76}:034007;
Voloshin M B. Phys. Rev. D (2007) {\bf 76}:014007;
Braaten E, Lu M. arXiv:0710.5482 [hep-ph] (2007).

\bibitem{maiani:2007}
Maiani L, Polosa A D, Riquer V. 
Phys. Rev. Lett. (2007) {\bf 99}:182003.

\bibitem{babar_y3940}
Aubert B {\etal} (BaBar Collaboration).
arXiv:0711.2047 [hep-ex] (2007) submitted to Physical Review Letters.

\bibitem{belle_x4160}
Pakhlov P {\etal} (Belle Collaboration).
arXiv:0708.3812 [hep-ex] (2007) submitted to Physical Review Letters.

\bibitem{belle_y4260}
Yuan C Z {\etal} (Belle Collaboration).
Phys. Rev. Lett. (2007) {\bf 99}:182004.

\bibitem{PDG} 
Yao W M {\etal} (Particle Data Group).
J. Phys. G (2006) {\bf 33}:1.

\bibitem{belle_y4660}
Wang X L {\etal} (Belle Collaboration).
Phys. Rev. Lett. (2007) {\bf 99}:142002.



\bibitem{belle_galina} Pakhlova G {\etal} (Belle Collaboration).
Phys. Rev. Lett. (2007) {\bf 98}:092001;
arXiv:0708.3313 [hep-ex] (2007) to appear in Physical Review Letters;
arXiv:0708.0082 (2007) [hep-ex] submitted to Physical Review D.

\bibitem{voloshin_rescattering}
Voloshin M B. arXiv:hep-ph/0602233 (2006).

\bibitem{mlyan_y4660} 
See, for example, 
Zhu S L. Phys. Lett. B (2005) {\bf 625}:212-216;
Close F E, Page P R. Phys. Lett. B (2005) {\bf 628}:215-222;
Kou E, Pene O. Phys. Lett. B (2005) {\bf 631}:164-169;
Ding~G J, Yan M-L. arXiv:0718.3712 [hep-ph] (2007).

\bibitem{belle_z4430}
Choi S-K {\etal} (Belle Collaboration)
arXiv:0708.1790 [hep-ex] (2007) submitted to Physical Review Letters.

\bibitem{molecule_z4430}
Rosner J L. Phys. Rev. D (2007) {\bf 76}:114002;
Meng C, Chao K-T.  arXiv:0708.4222 [hep-ph] (2007);
Braaten E, Lu M. arXiv:0712.3885 [hep-ph] (2007).

\bibitem{maiani:2007a}
Maiani L, Polosa A D, Riquer V. 
arXiv:0708.3997 [hep-ph] (2007).



\bibitem{babar_y2175}
Aubert B \etal (BaBar Collaboration). 
Phys. Rev. D (2006) {\bf 74}:091103.


\bibitem{ding_y2175}
Ding~G J, Yan M-L. 
Phys. Lett. B (2007) {\bf 650}:390-400
and Phys. Lett. B, 2007, {\bf 657}:49-54.

\bibitem{bes_y2175}
Ablikim M \etal (BESII Collaboration). 
arXiv:0712.1143 [hep-ex] submitted to Physical Review Letters.


\bibitem{babar_4s2pipi1s}
Aubert B \etal (BaBar Collaboration). 
Phys. Rev. Lett. (2006) {\bf 96}:232001.

\bibitem{belle_4s2pipi1s}
Sokolov A \etal (Belle Collaboration). 
Phys. Rev. D (2007) {\bf 75}:0710013.

\bibitem{belle_5s2pipi1s}
Chen K F \etal (Belle Collaboration). 
arXiv:0710.2577 [hep-ex] submitted to Physical Review Letters.

\bibitem{hou_yb}
Hou W-S. 
Phys. Rev. D (2006) {\bf 74}:017504.



\end{thebibliography}
\end{document}